\documentclass[conference]{IEEEtran}
\IEEEoverridecommandlockouts
\usepackage{cite}
\usepackage{amsmath,amssymb,amsfonts}
\usepackage{algorithmic}
\usepackage{graphicx}
\usepackage{textcomp}
\usepackage{xcolor}
\usepackage{url}
\def\BibTeX{{\rm B\kern-.05em{\sc i\kern-.025em b}\kern-.08em
    T\kern-.1667em\lower.7ex\hbox{E}\kern-.125emX}}
\begin{document}

\title{Connected and Exposed: Cybersecurity Risks, Regulatory Gaps, and Public Perception in Internet-Connected Vehicles\\
}

\author{\IEEEauthorblockN{Henrietta Hegyi}
\IEEEauthorblockA{\textit{Doctoral School on Safety and Security Sciences} \\
\textit{Obuda University}\\
Budapest, Hungary \\
hegyi.henrietta@uni-obuda.hu}
\and
\IEEEauthorblockN{L{\'a}szl{\'o} Erd{\H o}di}
\IEEEauthorblockA{\textit{Department of Informatics} \\
\textit{University of Oslo}\\
Oslo, Norway \\
laszloe@ifi.uio.no}

}

\maketitle

\begin{abstract}
The rapid advancement of Internet-connected vehicle technologies has introduced a new era of smart mobility, while simultaneously raising significant cybersecurity and privacy concerns. This paper explores the evolving threat landscape associated with connected vehicles, focusing on risks such as unauthorized remote access and the potential leakage of personal data. To assess the current state of protection, we conducted a comprehensive analysis of 16 international standards and regulations, evaluating them from multiple perspectives including regulatory strength, technical specificity, treatment of supply chain risks, and approaches to personal data handling.

In parallel, we carried out a user-focused survey designed to map consumer attitudes toward smart cars. The survey investigated which types of vehicles users trust and prefer, the reasons behind rejecting certain car types, their awareness of data-related risks, and whether they feel adequately informed about how their vehicles handle data. By combining regulatory analysis with user perception insights, this study aims to contribute to a more holistic understanding of the challenges and expectations surrounding connected vehicle ecosystems.
\end{abstract}


\section{Introduction}
The global adoption of Internet-connected vehicles (ICVs) has accelerated rapidly in recent years, transforming the automotive industry into a key frontier of the Internet of Things (IoT). Modern cars are increasingly equipped with advanced connectivity features such as real-time navigation, remote diagnostics and over-the-air updates. As manufacturers race to integrate smart technologies and autonomous capabilities, the number of connected vehicles on the road is projected to reach hundreds of millions globally within the next few years \cite{b1}. This widespread integration marks a shift in the role of cars—from isolated mechanical systems to dynamic, data-driven platforms—making them central to future mobility ecosystems. However, this growing connectivity also amplifies the potential attack surface, introducing new risks that impact not only driver safety and privacy but also broader critical infrastructure. As such, the significance of connected vehicles extends far beyond convenience, highlighting urgent needs for robust cybersecurity, international standards, and public awareness.

The functionality of ICVs evolves continously, expanding far beyond basic telematics into complex, data-rich systems \cite{b2}. These vehicles combine traditional automotive engineering with networked digital infrastructure, enabling both convenience and risk. To understand the cybersecurity implications of ICVs, it is essential to distinguish between their typical features—those that are now widely deployed across models—and non-typical or emerging features that are either experimental, regionally specific, or represent the next frontier in vehicle connectivity. Typical features of Internet-connected vehicles encompass a range of functionalities designed to enhance convenience, safety, and performance. Most modern models are equipped with telematics systems that provide real-time GPS navigation, vehicle tracking, and cloud-based route optimization. Infotainment systems are now standard, integrating with smartphones through Bluetooth, Apple CarPlay, or Android Auto to enable media streaming, voice assistants, and hands-free communication. Remote access features, often controlled via mobile apps, allow users to start or stop the engine, lock or unlock doors, locate their vehicle, and receive maintenance or diagnostic alerts. Over-the-air (OTA) updates have become a key feature, enabling manufacturers to deploy software patches and performance enhancements remotely, including updates to critical vehicle systems. Additionally, many vehicles include emergency response functions such as eCall, which automatically contacts emergency services with location data in the event of a crash. These core features define the baseline for connected vehicles today and serve as both value drivers and potential cybersecurity entry points.

As Internet-connected cars continue to evolve with rapid technological advancements, it is essential that standards, regulations, and public awareness keep pace with the emerging challenges. New features and increased connectivity introduce a broader and more complex threat landscape, requiring not only robust technical safeguards but also adaptive legal and regulatory frameworks. At the same time, the general public must be made aware of the potential cybersecurity risks associated with connected vehicles, as user behavior and trust are critical factors in the adoption and safe use of these technologies. This paper aims to map the current threat landscape facing connected cars and critically examine how existing standards and regulations respond to these risks. Additionally, we analyze the results of a public survey that explores how individuals perceive these threats, whether cybersecurity concerns influence their purchasing decisions, and how awareness of such risks may shape consumer attitudes toward certain countries of origin or vehicle manufacturers.

\section{Background}

\subsection{The threat landscape}

Internet-connected vehicles have been involved in several cybersecurity breaches and incidents. Some of these are related to remote or physical access, while others exploited vulnerabilities in firmware or third-party software components. The goals of the attacks also varied, ranging from taking control of the vehicle to compromising private data. 

\textit{Remotely accessing a vehicle} in recent years has been demonstrated by several cyber incidents and experiments. Hackers have exploited vulnerabilities in infotainment systems, telematics units, and wireless interfaces to gain unauthorized control over critical vehicle functions such as braking, steering, and acceleration—sometimes from miles away. Notable examples include the 2015 Jeep Cherokee hack \cite{b3}, where researchers remotely disabled the vehicle on a highway via a vulnerability in the \textit{Uconnect} system. These incidents highlight the growing attack surface introduced by connectivity and underscore the urgent need for secure software design, regular updates, and strict access controls in modern vehicles.

\textit{Accessing stored data} is also a significant cybersecurity concern, as these vehicles collect and retain large amounts of sensitive information. This data can include GPS history, contact lists, call logs, driving behavior, and even synced messages from mobile devices. In one notable case from 2022 \cite{b4}, security researchers were able to remotely access internal systems of multiple manufacturers via poorly secured APIs, retrieving location history and personal data from thousands of vehicles across different brands. Such incidents highlight serious privacy risks, particularly when personal or location data is accessed without consent, and they underscore the urgent need for stronger authentication, encryption, and data protection measures within the automotive ecosystem. 

\textit{Attacks targeting private data} have become increasingly common as well as cars store and transmit sensitive user information. Threat actors may exploit vulnerabilities in cloud-connected services, mobile apps, or vehicle infotainment systems to access data such as location history, contact lists, or even synced messages. In a substantial example from late 2024, a major data leak involving Volkswagen's software subsidiary Cariad exposed precise GPS location data for approximately 800,000 electric vehicles across brands like VW, Audi, Škoda, and SEAT \cite{b5}. The leaked data included real-time and historical location information, raising serious concerns about user tracking, privacy violations, and the potential misuse of behavioral data by malicious actors. This underscores the urgent need for secure authentication, encrypted communication, and strict access control mechanisms in the automotive ecosystem.

\textit{Attacks exploiting sensor data} pose also a growing threat as modern cars rely heavily on data from cameras, radar, LiDAR, GPS, and other onboard sensors to support advanced driver assistance systems (ADAS) and autonomous functions. Malicious actors can manipulate or spoof this data to mislead vehicle systems — for example, causing a car to misinterpret road signs or traffic conditions. A well-known case from 2020 demonstrated how researchers fooled a Tesla’s Autopilot system by placing small, strategically positioned stickers on a road sign, tricking the vehicle into accelerating to 85 mph in a 35 mph zone \cite{b6}. Such attacks highlight the vulnerability of sensor-driven systems and the need for sensor fusion validation and anomaly detection to ensure data integrity and safety.

\textit{Attacks targeting the CAN bus} pose another serious risk to ICVs, as the Controller Area Network (CAN) is the primary internal communication channel between critical vehicle components such as the engine control unit, brakes, and steering. By injecting malicious messages into the CAN bus—often via compromised telematics units, OBD-II ports, or Bluetooth interfaces—attackers can manipulate vehicle behavior with precision. In the previously mentioned Jeep Cherokee hack case \cite{b3} the attackers did lateral movement using the CAN bus. After compromising \textit{Uconnect}, they pivoted deeper into the vehicle’s internal networks and reached the CAN bus — the vehicle’s core communication backbone. By exploiting the CAN bus, they succeeded to disable brakes, cut the engine, and control steering—all from over a mile away. This incident starkly highlights the potential for remote CAN bus attacks to compromise safety-critical systems and underscores the need for network segmentation, message authentication, and intrusion detection tools within vehicles.

\textit{Firmware attacks} target the low-level software that controls hardware components such as ECUs (Electronic Control Units), infotainment systems, or telematics modules. Attackers may exploit vulnerabilities in firmware update mechanisms to install malicious code, either over physical interfaces or wirelessly via compromised update servers or spoofed over-the-air updates. For example, in 2018, researchers from Tencent’s Keen Security Lab demonstrated a full chain of attacks on a Tesla Model S \cite{b7}, where they exploited vulnerabilities in the browser and the firmware of the gateway ECU. This allowed them to remotely control vehicle functions such as opening doors or applying brakes, even while the car was in motion. Firmware attacks are particularly dangerous because they can persist across reboots and are difficult for users to detect—highlighting the critical need for secure boot processes, digitally signed firmware, and protected update channels.

\textit{Supply chain attacks} exploit vulnerabilities in third-party components or services—such as telematics dongles, infotainment modules, or backend software—to inject malicious code into the vehicle ecosystem. These attacks often occur before a vehicle is even on the road, compromising hardware or software updates during manufacturing, distribution, or integration phases. A well-known example is the 2015 case where researchers at UC San Diego exploited insecure OBD-II dongles—used by insurance firms and fleet operators—to remotely inject malicious messages into the CAN bus of a 2013 Corvette \cite{b8}. They were able to manipulate critical functions like windshield wipers and brakes via SMS commands to the dongle, illustrating how a seemingly benign third-party accessory can serve as a backdoor into vehicle control systems. This highlights the urgent need for rigorous vetting, secure update mechanisms, and supply chain transparency in automotive cybersecurity.

\subsection{Standards and regulation}
The rapid evolution of ICVs has prompted a growing ecosystem of cybersecurity, safety, and data protection standards aimed at mitigating the complex risks introduced by increased digitalization. A range of international frameworks and regulations now govern different aspects of ICV development and operation—from cybersecurity engineering processes and software update management to functional safety and consumer data privacy. 

General cybersecurity standards such as ISO/IEC 27001 \cite{b9} and the NIST Cybersecurity Framework \cite{b10} provide foundational principles for managing information security and risk across any organization, including automotive manufacturers. Building on this, ISO/IEC 15408 (Common Criteria) \cite{b11} establishes a framework for evaluating the security of IT products, while TISAX (Trusted Information Security Assessment Exchange) \cite{b12} adapts ISO/IEC 27001 specifically to the automotive industry’s needs for secure data exchange and third-party assessments.

In the automotive context, ISO/SAE 21434 \cite{b13} is the principal standard for cybersecurity risk management throughout the vehicle lifecycle, addressing threats from design to decommissioning. Complementing this, UNECE WP.29 regulations R155 \cite{b14} and R156 \cite{b15} make cybersecurity and software update management mandatory for type approval in many countries, effectively enforcing ISO/SAE 21434 principles in law. Supporting the safety side, ISO 26262 \cite{b16} defines functional safety requirements for electronic systems, while ISO 21448 (SOTIF – Safety of the Intended Functionality) \cite{b17} expands the scope to hazards arising from system limitations or misuse.

To ensure security in vehicle software development, SAE J3061 \cite{b18} provides a cybersecurity process framework that closely aligns with ISO/SAE 21434, while ASPICE (Automotive SPICE) \cite{b19} focuses on the maturity and quality of automotive software processes. The AUTOSAR Adaptive Platform \cite{b20} further supports cybersecurity by offering standardized architectures and services designed for high-performance computing and connected ECUs.

With the increasing focus on software and over-the-air updates, ISO 24089 \cite{b21} addresses software update engineering for road vehicles, ensuring secure and verifiable update processes. Electromagnetic compatibility is handled by CISPR 25 \cite{b22}, which limits radio interference from electrical/electronic components. From a privacy perspective, GDPR \cite{b23} sets legal requirements for the handling of personal data, which is highly relevant as connected vehicles increasingly collect, transmit, and store user information.

The Cyber Resilience Act (CRA) \cite{b24} proposed by the EU introduces cybersecurity requirements for a wide range of digital products, including components used in connected vehicles. Finally, in the U.S., the NHTSA Cybersecurity Guidelines \cite{b25} provide voluntary guidance to help manufacturers manage evolving threats and implement best practices for vehicle cybersecurity.

\subsection{Related work}
Recent research has extensively explored the cybersecurity challenges faced by ICVs, highlighting both the breadth of threats and the evolving landscape of defense mechanisms. Comprehensive surveys such as those by Chowdhury et al. \cite{b26} and Pham \& Xiong \cite{b27} map out the range of vulnerabilities in connected and autonomous vehicles (CAVs), from network attacks to ECU-level exploits, while also identifying gaps between theoretical models and practical deployment. Annabi et al. \cite{b28} further advance this by introducing the Zero Trust model as a paradigm shift, emphasizing continuous verification across V2V and V2I environments to prevent lateral movement and privilege escalation. Specific attack vectors such as those targeting in-vehicle networks are addressed in works by Pavlenko et al. \cite{b29} and Liu \& Yang \cite{b30}, who demonstrate how weaknesses in the CAN bus can be exploited to compromise braking systems or inject unauthorized messages—underscoring the urgency of implementing message authentication and ECU integrity verification. Technical solutions to these issues are further discussed by researchers such as Marksteiner \& Ma \cite{b31}, who advocate for the automation of cybersecurity testing in development pipelines. In the context of software architecture, the work by Razzaque et al. \cite{b32} explores vulnerabilities in vehicle-to-grid (V2G) systems, warning that bi-directional communication with power grids introduces additional threat vectors, particularly in EV ecosystems. Complementing these technical studies, Khan et al. \cite{b33} propose a modeling-based cybersecurity framework that incorporates technical, behavioral, and organizational variables, offering a more systemic approach to CAV security. Collectively, these studies underline the need for multilayered defenses, robust standards, and continuous assessment to protect the rapidly expanding attack surface of modern vehicles.
Beyond technical attack methods and defenses, a growing body of literature emphasizes the broader consequences of cybersecurity vulnerabilities in ICVs. These risks extend far beyond the immediate disruption of driving functionality, posing serious implications for privacy, surveillance, national security, and public trust. The study by López et al. \cite{b34} explores how modern passenger vehicles can serve as unintentional cyber threat sources, demonstrating the potential for smart vehicles to be repurposed for surveillance and intelligence-gathering—either by malicious actors or through legal yet ethically questionable practices. Such analyses highlight that vulnerabilities in ICVs can be exploited not only for control or data theft but also for persistent tracking, behavior monitoring, and cross-domain cyber-physical campaigns. Other studies \cite{b35} estimate the amount of data that Internet connected vehicles can send in covert channels if the vehicles are used for surveillance by threat actors. These insights underline the importance of securing both the technological and societal dimensions of vehicle connectivity.

While the aforementioned scientific papers (\cite{b26}–\cite{b33}) offer valuable insights into the technical vulnerabilities and emerging threats in Internet-connected vehicles, they largely focus on attack vectors, detection mechanisms, and mitigation strategies from an engineering and system-level perspective. However, these studies often do not address how existing standards and regulations respond to these challenges—if at all. There remains a noticeable gap in the literature regarding the alignment (or misalignment) between real-world threat scenarios and the countermeasures mandated or guided by standards such as ISO/SAE 21434, UNECE R155 and R156 or NIST's Cybersecurity Framework. This disconnect highlights the need for a more integrated view that bridges technical findings with regulatory frameworks to ensure a holistic cybersecurity posture for modern vehicles.

\section{Cyber threats and regulations}
The regulatory and standards landscape for Internet-connected vehicles is highly fragmented, reflecting the diverse origins, scopes, and enforcement mechanisms of the listed frameworks. As shown in the Table \ref{tab:standards_overview}., only a few standards—most notably UNECE R155 and R156, the Cyber Resilience Act (CRA), and GDPR—are mandatory and legally enforced, primarily within the European Union and affiliated UNECE countries. These impose tangible sanctions such as denial of market access or heavy financial penalties for non-compliance. In contrast, many of the other frameworks like ISO/SAE 21434, ISO 26262, and AUTOSAR Adaptive Platform are not mandatory but are increasingly adopted by OEMs and Tier-1 suppliers due to industry pressure and the need to meet contractual or reputational expectations. Some standards, such as ISO/IEC 15408 and TISAX, operate through certification processes that are often indirectly required in B2B environments, even if not enforced by law. Supervision also varies significantly: while GDPR and UNECE regulations are monitored by government bodies, others rely on consortiums (like AUTOSAR or ENX) or are self-imposed through audits and supplier evaluations. The lack of uniform global enforcement and the reliance on voluntary adoption for many cybersecurity standards create inconsistencies in how vehicle manufacturers address threats. This mixed ecosystem complicates efforts to ensure a universally secure baseline for connected vehicles, particularly in cross-border markets where differing regulatory regimes apply. Overall, while technical guidance is abundant, a stronger convergence between regulation, enforcement, and industrial adoption is needed to comprehensively address the cybersecurity risks of modern intelligent transportation systems.

\begin{table*}[!t]
\caption{Overview of Selected Standards and Regulations}
\centering
\begin{tabular}{|p{3.5cm}|p{3.2cm}|p{5.5cm}|p{3.5cm}|}
\hline
\textbf{Standard/Regulation} & \textbf{Mandatory?} & \textbf{Sanctions} & \textbf{Supervision} \\
\hline
ISO/SAE 21434 & Not mandatory & No official sanctions & No official supervision \\
\hline
UNECE WP.29 – R155, R156 & Yes, mandatory in the EU and UNECE countries & Type approval refusal, market exclusion & National transport authorities \\
\hline
ISO 26262 & Not mandatory everywhere & OEM may reject supplier product or contract & No direct official supervision \\
\hline
ISO/IEC 27001 & Not mandatory in the automotive industry & Lack of certification can have contractual consequences & Auditing based on the standard \\
\hline
NIST Cybersecurity Framework & Not mandatory & No legal sanctions, but reputational damage possible & No direct official supervision \\
\hline
AUTOSAR Adaptive Platform & Not mandatory & No official sanctions & AUTOSAR consortium \\
\hline
ISO/IEC 15408 (Common Criteria) & Not mandatory & Loss of market access without certification & National government certifications \\
\hline
ISO 24089 & Not mandatory & OEMs may require it, no legal sanctions & No direct supervision \\
\hline
CRA & Mandatory & European Commission, ENISA, national authorities & Yes, severe fines like GDPR \\
\hline
GDPR & Yes, mandatory in the EU & Fines (up to 4\% of global revenue) & EU member state data protection authorities \\
\hline
CISPR 25 & Generally mandatory & Market approval refusal & National authorities \\
\hline
TISAX & Not legally mandatory & Contractual exclusion possible & ENX Association \\
\hline
SAE J3061 & Not mandatory & May lead to competitive disadvantage & No official supervision \\
\hline
ASPICE & Not legally mandatory & OEM may exclude supplier & Automotive manufacturer audits \\
\hline
ISO 21448 & Not mandatory & No direct sanctions & No official supervision \\
\hline
NHTSA Cybersecurity Guidelines & Not mandatory & No official sanctions, reputational risks & NHTSA (USA) \\
\hline
\end{tabular}
\label{tab:standards_overview}
\end{table*}

We also carried out a comprehensive analysis of these standards and regulations considering the different threats. Our investigation revealed how they address the previously mentioned threat vectors in connected vehicles. These include remote access, data storage, sensor integrity, in-vehicle communication (e.g., CAN), firmware security, supply chain risks, and personal data protection. The analysis highlights not only the scope and depth of coverage but also the enforceability and practical implications of each framework.

One of the most striking findings is the divergence in regulatory strength and enforceability. Frameworks like UNECE R155, R156 and the Cyber Resilience Act (CRA) impose mandatory requirements, backed by regulatory oversight and homologation processes. These contrast with voluntary standards such as SAE J3061 or NIST CSF, which provide valuable guidance but lack legal force. This distinction is critical for manufacturers, as compliance with binding regulations is essential for market access, while voluntary frameworks serve more as internal benchmarks or tools for continuous improvement.

The technical depth and specificity of the standards also vary significantly. For example, ISO/SAE 21434 offers detailed technical requirements tailored to the automotive context, particularly in areas like firmware integrity and CAN bus risk analysis. In contrast, ISO/IEC 27001 and Common Criteria (ISO/IEC 15408) provide robust security principles but are not vehicle-specific, requiring adaptation for automotive use. This creates a fragmented landscape where manufacturers must integrate multiple standards to achieve comprehensive protection across all threat domains.

Another key insight is the inconsistent treatment of supply chain risks. While several standards acknowledge the importance of supplier involvement, only a few—such as CRA and UNECE R155, R156 —introduce mechanisms for accountability or transparency. Even then, enforcement often depends on the OEM’s leverage over suppliers, as many standards do not impose direct obligations on third parties. This gap poses a significant challenge, given the complexity and global nature of automotive supply chains.

The handling of personal data and privacy is also uneven. Only GDPR and, to some extent, ISO/IEC 27001 and TISAX, explicitly address data protection. Most automotive cybersecurity standards focus on system integrity and operational safety, often sidelining privacy concerns. This separation suggests that cybersecurity compliance alone is insufficient in jurisdictions with strong data protection laws, and manufacturers must ensure parallel alignment with privacy regulations.

The comparison also revealed that no single standard offers complete coverage across all threat categories. Each framework has its strengths and blind spots, shaped by its origin, scope, and intended audience. As a result, a layered approach—combining regulatory compliance, technical standards, and organizational best practices—is essential for building resilient and legally compliant connected vehicle systems.

\begin{table*}[!t]
\caption{Regulations and Threats}
\centering
\begin{tabular}{|p{1.75cm}|p{1.75cm}|p{1.75cm}|p{1.75cm}|p{1.75cm}|p{1.75cm}|p{1.75cm}|p{1.75cm}|}
\hline
\textbf{Standard/} & \textbf{Remote access} & \textbf{Stored data} & \textbf{Sensor} & \textbf{CAN} & \textbf{Firmware}& \textbf{Supply chain} & \textbf{Private data}\\
\textbf{Regulation} & & & & & & & \\
\hline

\textbf{ISO/SAE 21434} & \checkmark Technical vehicle-focused requirements, not legally enforced & \checkmark Part of risk analysis; handles user data indirectly & $\triangle$ Sensor fusion threats may be required, but no deep analysis & $\triangle$ No protocol-level protection, but network risk analysis expected & \checkmark Covers software integrity, not in deep technical detail & $\triangle$ Supplier audits expected, but no enforcement authority & $\times$ Not a primary goal; does not focus on user data protection \\ 
\hline

\textbf{UNECE WP.29 – R155, R156} & \checkmark CSMS is mandatory; enforced via type approval & \checkmark Protects data integrity/ confidentiality, but not a privacy law & $\triangle$ Limited focus; mainly addresses software/network risks & \checkmark CAN protection required under R155 & \checkmark Software update management mandatory; legally enforced & $\triangle$ Supplier involvement in CSMS required, but data sharing not enforceable & $\times$ Not a privacy regulation; does not replace GDPR \\
\hline

\textbf{ISO26262} & $\times$ Focuses on functional safety, not cyberattacks. & $\times$ Not addressed. & \checkmark Sensor and fusion fault analysis mandatory. Good coverage. & $\triangle$ Handles comms errors, not active attacks. &
$\triangle$ Analyzes software faults, not hostile tampering. & $\times$ Suppliers not evaluated from security perspective. & $\times$ User protection not a goal. \\
\hline

\textbf{ISO/IEC 27001} & \checkmark Regulates IT access control, not vehicle-specific. & \checkmark Detailed on data protection, encryption, access. & $ \times $ Not applicable to physical sensors. & $ \times $ Not automotive-specific. & $ \triangle $ May include server-side firmware handling, not vehicle-specific. & $ \triangle $ Supplier risk handling depends on supplier cooperation. & \checkmark Protects personal and business data, enterprise-focused. \\
\hline 

\textbf{NIST Cybersecurity Framework} & \checkmark Umbrella framework, not industry-specific but useful. &
\checkmark Includes protect, detect, recover functions. & $ \times $ Not addressed. & $ \times $ Not directly relevant. & $ \triangle $ Firmware may fall under 'Identify' and 'Protect'. & $ \triangle $ Supplier risk depends on supplier cooperation. & \checkmark Includes privacy aspects, not legally binding. \\
\hline

\textbf{AUTOSAR Adaptive Platform} & $ \triangle $ Provides tech foundation, but no protection rules. & $ \times $ Not addressed. & $ \times $ Not addressed. & $ \times $ Not addressed. & \checkmark Supports secure implementation (architecture/dev). & \checkmark Supports secure implementation (architecture/dev). & $ \times $ Not addressed. \\
\hline

\textbf{ISO/IEC 15408 (Common Criteria)} & \checkmark Evaluated via security goals and assurance levels – detailed, not vehicle-specific. &
\checkmark Evaluated via security goals and assurance levels – detailed, not vehicle-specific. & $\times$ Not relevant in this context. & $\times$ Not relevant in this context. & $\times$ Not relevant in this context. & $\times$ Not relevant in this context. & \checkmark Evaluated via security goals and assurance levels – detailed, not vehicle-specific. \\
\hline

\textbf{ISO 24089} & $\times$ Focuses solely on software updates. & $\times$ Focuses solely on software updates. & $\times$ Focuses solely on software updates. & $\times$ Focuses solely on software updates. & \checkmark Regulates software update lifecycle – detailed but not full protection. & $\times$ Focuses solely on software updates. & $\times$ Focuses solely on software updates. \\
\hline

\textbf{CRA} & \checkmark Strong regulation of network-connected devices incl. remote access; security updates mandatory. &
\checkmark Devices must protect stored data (e.g., config, logs). & $\triangle$ Sensors not named, but applies to all connected digital components. & $\times$ Not for internal vehicle bus protocols (e.g., CAN), unless they qualify as standalone digital products. & \checkmark Regulates secure firmware development, updates, and support. & $\triangle$ Requires supply chain transparency and component security; indirect responsibility for suppliers. & $\times$ Not focused on data protection—covered by GDPR; applies if security flaws cause data compromise. \\
\hline

\textbf{GDPR} & $\triangle$ Applies only if personal data is accessed. & $\triangle$ Covers legal basis and security expectations, not all data types. & $\times$ Not technical; sensor data only if personal. & $\times$ Not applicable. & $\times$ Not addressed. & $\triangle$ Regulates responsibility in joint data control; only from privacy perspective. & \checkmark Specifically aims to protect personal data; legally enforceable. \\
\hline

\textbf{CISPR 25} & $\times$ Focuses only on electromagnetic compatibility. & $\times$ Focuses only on electromagnetic compatibility. & $\times$ Focuses only on electromagnetic compatibility. & $\times$ Focuses only on electromagnetic compatibility. & $\times$ Focuses only on electromagnetic compatibility. & $\times$ Focuses only on electromagnetic compatibility. & $\times$ Focuses only on electromagnetic compatibility. \\
\hline



\end{tabular}
\label{tab:standards_and_threats}
\end{table*}

\begin{table*}[!t]
\centering
\begin{tabular}{|p{1.75cm}|p{1.75cm}|p{1.75cm}|p{1.75cm}|p{1.75cm}|p{1.75cm}|p{1.75cm}|p{1.75cm}|}
\hline
\textbf{Standard/} & \textbf{Remote access} & \textbf{Stored data} & \textbf{Sensor} & \textbf{CAN} & \textbf{Firmware}& \textbf{Supply chain} & \textbf{Private data}\\
\textbf{Regulation} & & & & & & & \\
\hline

\textbf{TISAX} & $\times$ Not addressed. & \checkmark Audit framework for data security – not technical protection. Corporate level. & $\times$ Not addressed. & $\times$ Not addressed. & $\times$ Not addressed. & \checkmark Audit framework for data security – not technical protection. Corporate level. & \checkmark Audit framework for data security – not technical protection. Corporate level. \\
\hline

\textbf{SAE J3061} & \checkmark Recommends good practices, not mandatory. No official oversight & $\triangle$ Mentioned, but not detailed & $\triangle$ Mentioned, but not detailed & $\triangle$ Mentioned, but not detailed & \checkmark Recommends good practices, not mandatory. No official oversight & \checkmark Recommends good practices, not mandatory. No official oversight & $\triangle$ Mentioned, but not detailed \\
\hline

\textbf{ASPICE} & $\triangle$ Provides tech basis, but doesn’t regulate protection mechanisms & $\times$ Does not cover this area & $\times$ Does not cover this area & $\times$ Does not cover this area & \checkmark Supports secure implementation at architecture and development level & \checkmark Supports secure implementation at architecture and development level & $\times$ Does not cover this area \\
\hline

\textbf{ISO 21448} & $\times$ Not intended to address cyber threats & $\times$ Not intended to address cyber threats & \checkmark Focuses on sensor faults in functional error analysis. Functionally detailed  & $\times$ Not intended to address cyber threats & $\times$ Not intended to address cyber threats & $\times$ Not intended to address cyber threats & $\times$ Not intended to address cyber threats \\
\hline
\textbf{NHTSA Cybersecurity Guidelines} & $\triangle$ Strong recommendations, not legally binding. Advised for U.S. manufacturers & $\times$ Touches on it, but not mandatory & $\times$ Touches on it, but not mandatory & $\triangle$ Strong recommendations, not legally binding. Advised for U.S. manufacturers & $\triangle$ Strong recommendations, not legally binding. Advised for U.S. manufacturers & $\triangle$ Strong recommendations, not legally binding. Advised for U.S. manufacturers & $\times$ Touches on it, but not mandatory \\
\hline

\end{tabular}
\label{tab:standards_and_threats2}
\end{table*}

\section{Public Attitudes Toward Cybersecurity risks}

As part of our research, we conducted a user survey to explore public behavior and acceptance regarding the cyber security risks associated with internet-connected vehicles. Around 300 responses were received, mainly from Europe, but approximately 10 percent of the respondents were from outside Europe. The survey aimed to map how respondends consider security related concerns when buying a new vehicle. The reason of rejecting brands or cars from particular countries was also analyzed as well as the information the respondents received from the vendors and manufacturers. By analyzing user attitudes and expectations, we sought to complement the technical comparison of standards with insights into real-world acceptance and potential barriers to adoption of cyber security measures.

The survey consisted of some background info questions such as the number of cars the participants have or the age of the car they have. In addition, several open ended questions were also part of the survey such as \textit{What car is considered as a smart car?} or \textit{Could it be problematic if the vehicle transmits data to the manufacturer or 3rd parties without the users' knowledge? }. The survey question are presented as well as the type of the questions in Table \ref{tab:survey_questions}.

\begin{table*}[ht]
\centering
\begin{tabular}{|p{9cm}|p{5cm}|}
\hline
\textbf{Survey question} & \textbf{Answer Type} \\
\hline
Which country are you from? & List of countries \\
\hline
Which age group do you belong to? & List of age intervals \\
\hline
Do you currently own a personal vehicle? & Yes/No \\
\hline
If no, have you ever owned a car? & Yes/No \\
\hline
If yes, which brand and type of car do you have or had last time? & List of car brands \\
\hline
How old is (was) your car? & List of age intervals \\
\hline
Do you know that your car can connect to the Internet? & Yes/No \\
\hline
How many cars do you own in your immediate family (parents, siblings, children)? & List of age intervals \\
\hline
What do you think, what car is considered a "smart car"? & Open ended \\
\hline
Is/was your car a smart car? & Yes/No \\
\hline
Is there a smart car in your family? & Yes/No \\
\hline
Do you think your car sends (sent) data to the manufacturer? & Yes/No \\
\hline
Do you think your car sends (sent) data to third party companies? & Yes/No \\
\hline
Do you think the cars of your family members send data to the manufacturer or other companies? & Yes/No \\
\hline
Do you think it is important for the user to be informed if his/her/their car sends data to the manufacturer or to data analysis companies? & Yes/No \\
\hline
Have you ever received any information about what data your car is sending? If so, in what form? (e.g. data protection information, service, brand store) & Yes/No \\
\hline
Do you think it could cause a problem, and what kind of problem, if the vehicle transmits data to the manufacturer or third parties without the user's knowledge? & Open ended \\
\hline
When choosing a car, do you have a preference in terms of country of origin? Which country is it? & Open ended \\
\hline
Is there a country you wouldn't choose a car from? If so, why? If not, why not? & Open ended \\
\hline
\end{tabular}
\caption{Survey Questions}
\label{tab:survey_questions}
\end{table*}

\begin{figure}[htbp]
\centerline{\includegraphics[width=\columnwidth]{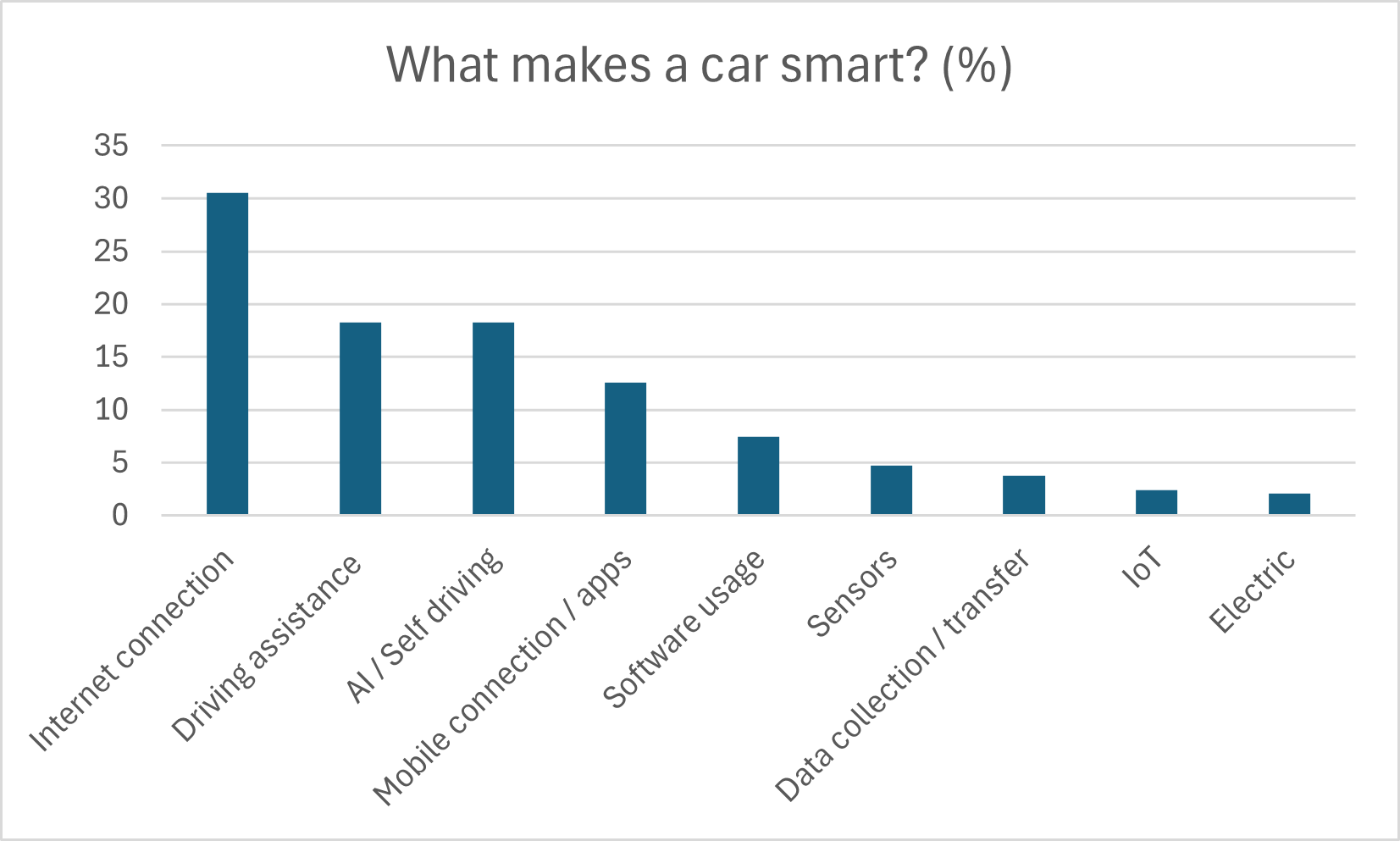}}
\caption{What makes a car smart?}
\label{fig:smart}
\end{figure}

The term “smart car” is inherently ambiguous, as it lacks a universally accepted definition. Different individuals associate varying features or technologies with the concept—some emphasize connectivity, others prioritize automation or driver assistance systems. This ambiguity is reflected in the survey results shown in Figure \ref{fig:smart}, where participants highlighted diverse aspects they consider central to the idea of a smart car. The range of answers underscores how the term functions more as a general label for technological advancement in vehicles than as a precise descriptor. This suggests that while flashy features like autonomous driving or infotainment receive significant attention, underlying concerns such as cybersecurity and data privacy tend to be overlooked or underestimated by the general public. This gap highlights a potential blind spot in user awareness, where security is relegated to the background despite its critical importance.

The survey results reveal a clear regional preference among respondents, most of whom were based in Europe. When asked which countries' cars they prefer, the majority indicated a strong inclination toward European brands—particularly highlighting German manufacturers—as well as Japanese vehicles. Although the question was framed to elicit positive preferences, the open-ended nature of the responses led a few participants to express their choices in negative terms, stating they preferred “anything but Chinese or American cars.” This pattern suggests that regional loyalty and perceived quality or trustworthiness may significantly influence consumer attitudes in the automotive sector, especially in relation to connected or smart vehicles.

\begin{figure}[htbp]
\centerline{\includegraphics[width=\columnwidth]{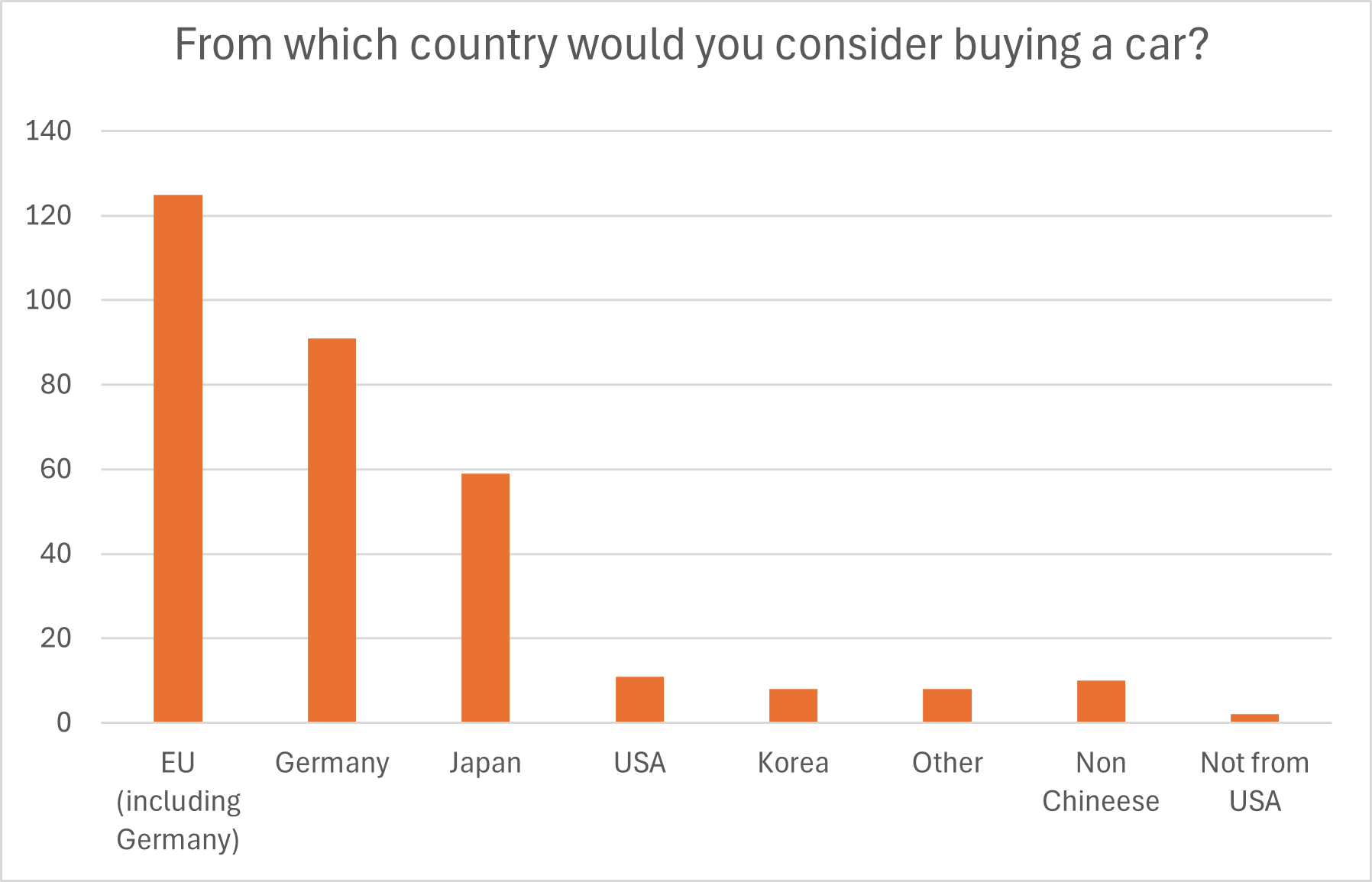}}
\caption{From which country would you consider buying a car?}
\label{fig:preference}
\end{figure}

The questionnaire also asked respondents to indicate negative preferences, revealing that a significant majority expressed concerns about Chinese cars. It is important to note, however, that the question was framed in a way that required participants to name at least one country whose vehicles they would be least likely to choose, even if they did not hold strong negative views. Notably, among the predominantly European respondents, Japanese cars were rarely rejected, indicating a generally positive perception and low level of distrust toward Japanese automotive brands in this demographic.

\begin{figure}[htbp]
\centerline{\includegraphics[width=\columnwidth]{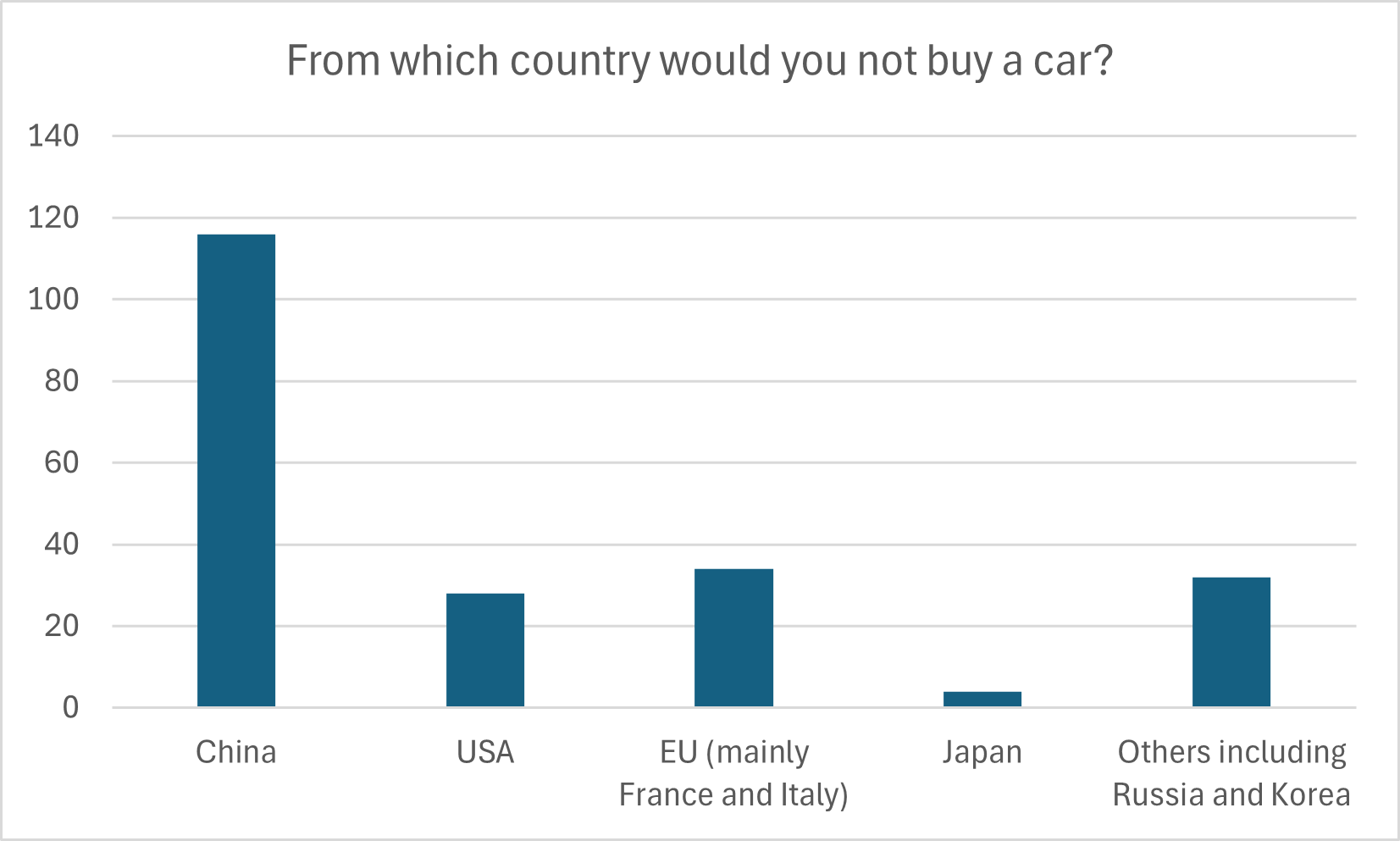}}
\caption{From which country would you not buy a car?}
\label{fig:negativepreference}
\end{figure}

\begin{figure}[htbp]
\centerline{\includegraphics[width=\columnwidth]{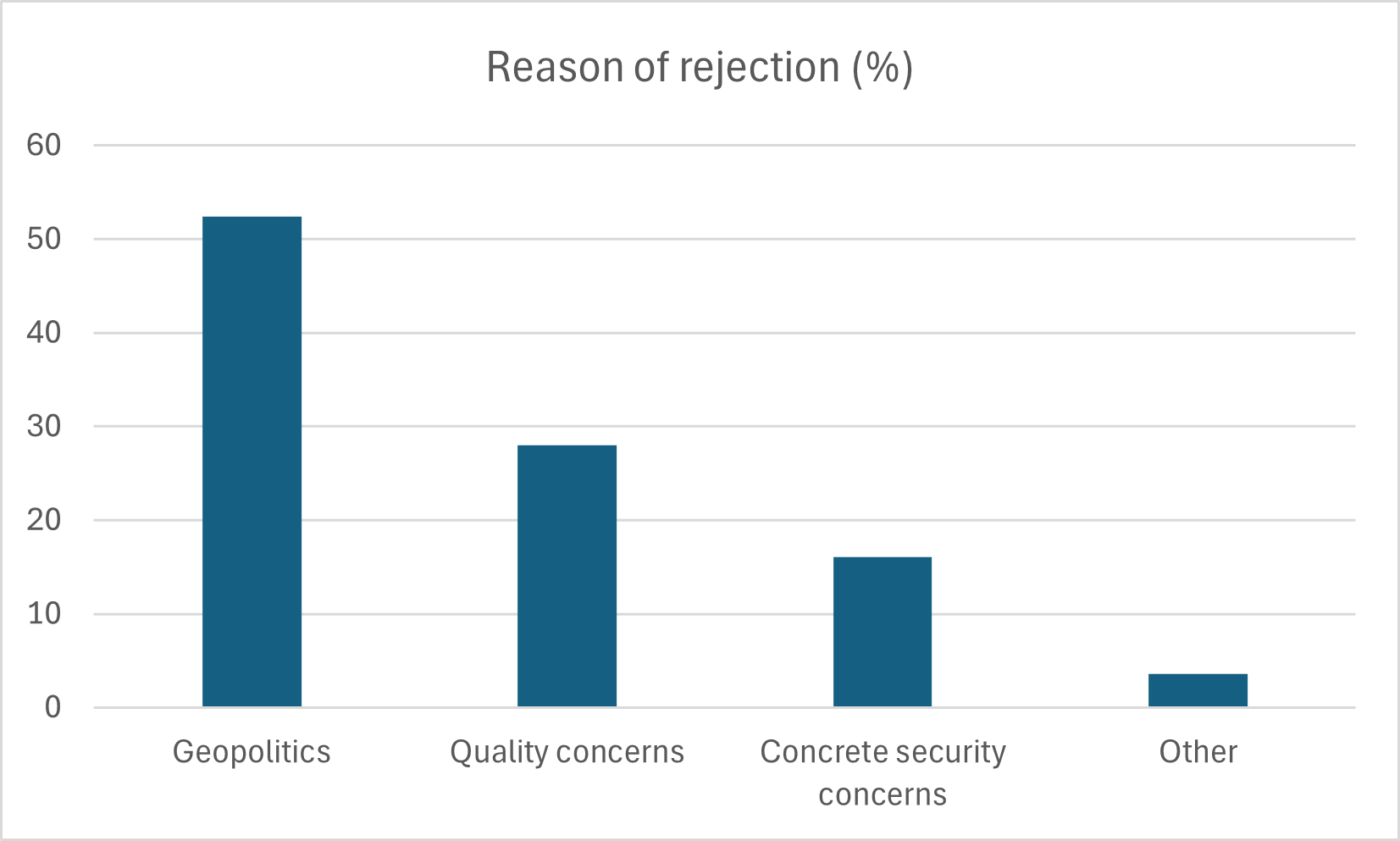}}
\caption{Reason of rejection}
\label{fig:rejection}
\end{figure}

The survey also included an open-ended question aimed at understanding the reasons behind participants' negative preferences toward certain vehicles. This allowed respondents to elaborate in detail on why they distrust specific types of cars. Based on the responses, we categorized the justifications into three main groups: geopolitical concerns, perceived quality issues, and concrete safety concerns. Geopolitical and safety-related concerns often appeared together; however, we made efforts to distinguish whether the distrust stemmed from a specific technical or safety issue, or if it was solely based on the vehicle's country of origin. A significant portion of respondents cited geopolitical reasons as their primary concern (Fig \ref{fig:rejection}.).

\begin{figure}[htbp]
\centerline{\includegraphics[width=\columnwidth]{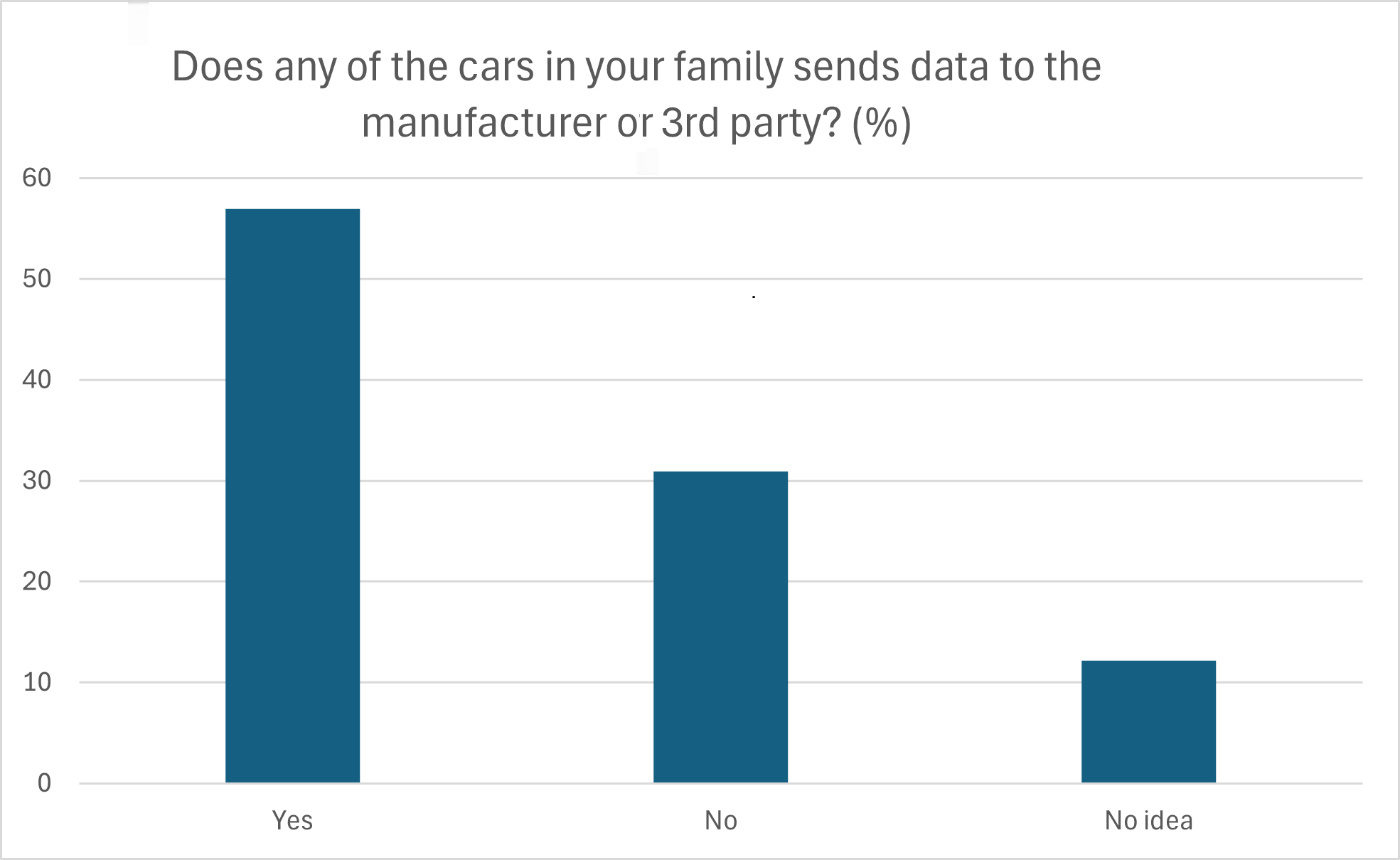}}
\caption{Percentage of respondents for the data sending answers}
\label{fig:datatransfer}
\end{figure}

The responses to the survey indicate that nearly 60\% of participants believe their vehicles transmit data both to the manufacturers and to third-party entities. This proportion is even significantly higher among respondents who own cars less than five years old. These results suggest that car owners are increasingly aware of the possibility that their vehicles may be collecting and sharing data, especially in the case of modern, connected vehicles (Fig. \ref{fig:datatransfer}.)

\begin{figure}[htbp]
\centerline{\includegraphics[width=\columnwidth]{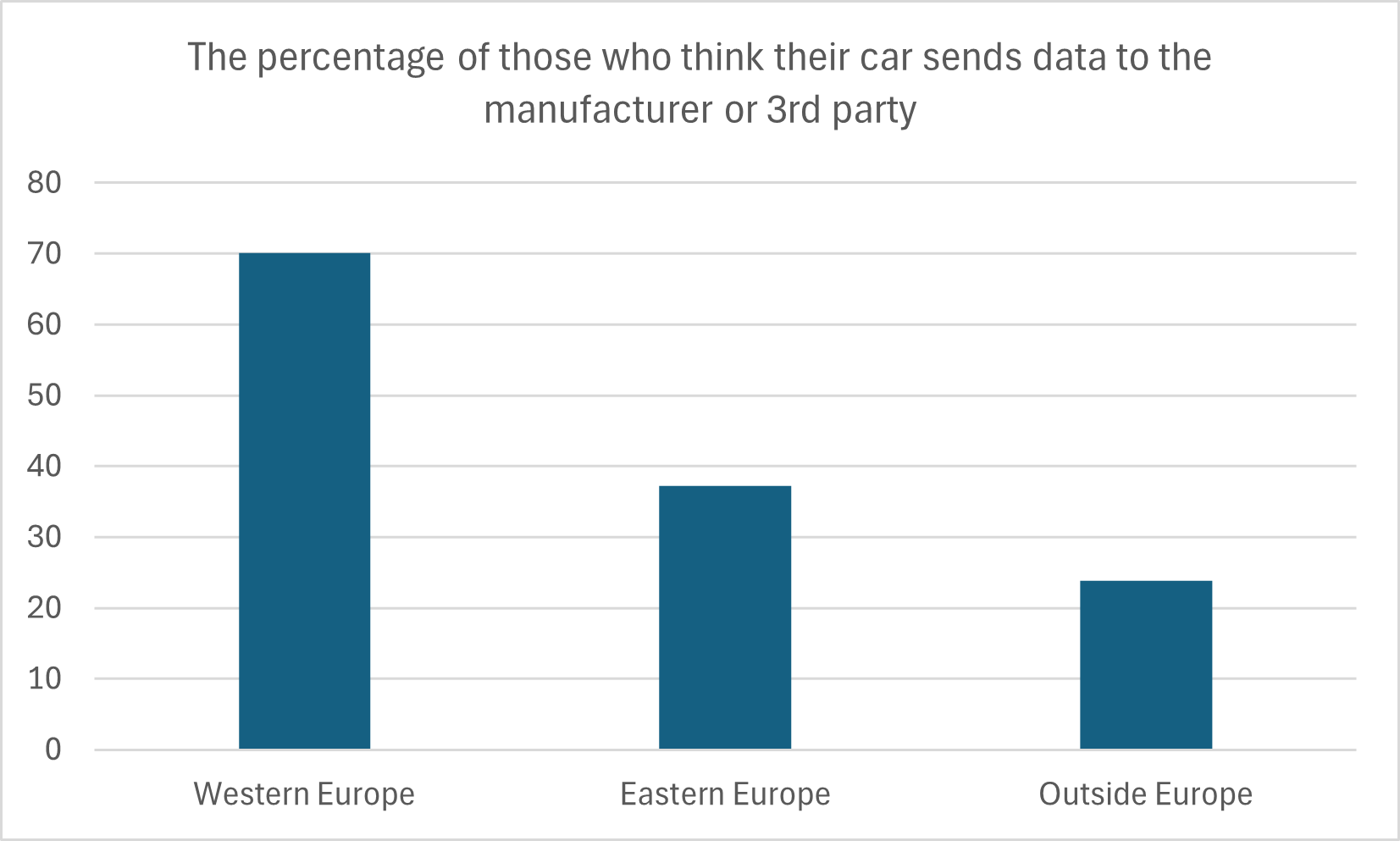}}
\caption{Percentage of respondents who think their car sends data}
\label{fig:datatransferpercentage}
\end{figure}

The survey also reveals a clear regional difference in perceptions: respondents from Western Europe are much more likely to believe that their cars are sending data to various destinations. In contrast, this awareness appears to be lower among Eastern European respondents. While we also collected responses from outside Europe, the geographical distribution of these participants was uneven—mostly from South America—and the number of responses from these regions was significantly lower, limiting broader conclusions (Fig. \ref{fig:datatransferpercentage}).

\begin{figure}[htbp]
\centerline{\includegraphics[width=\columnwidth]{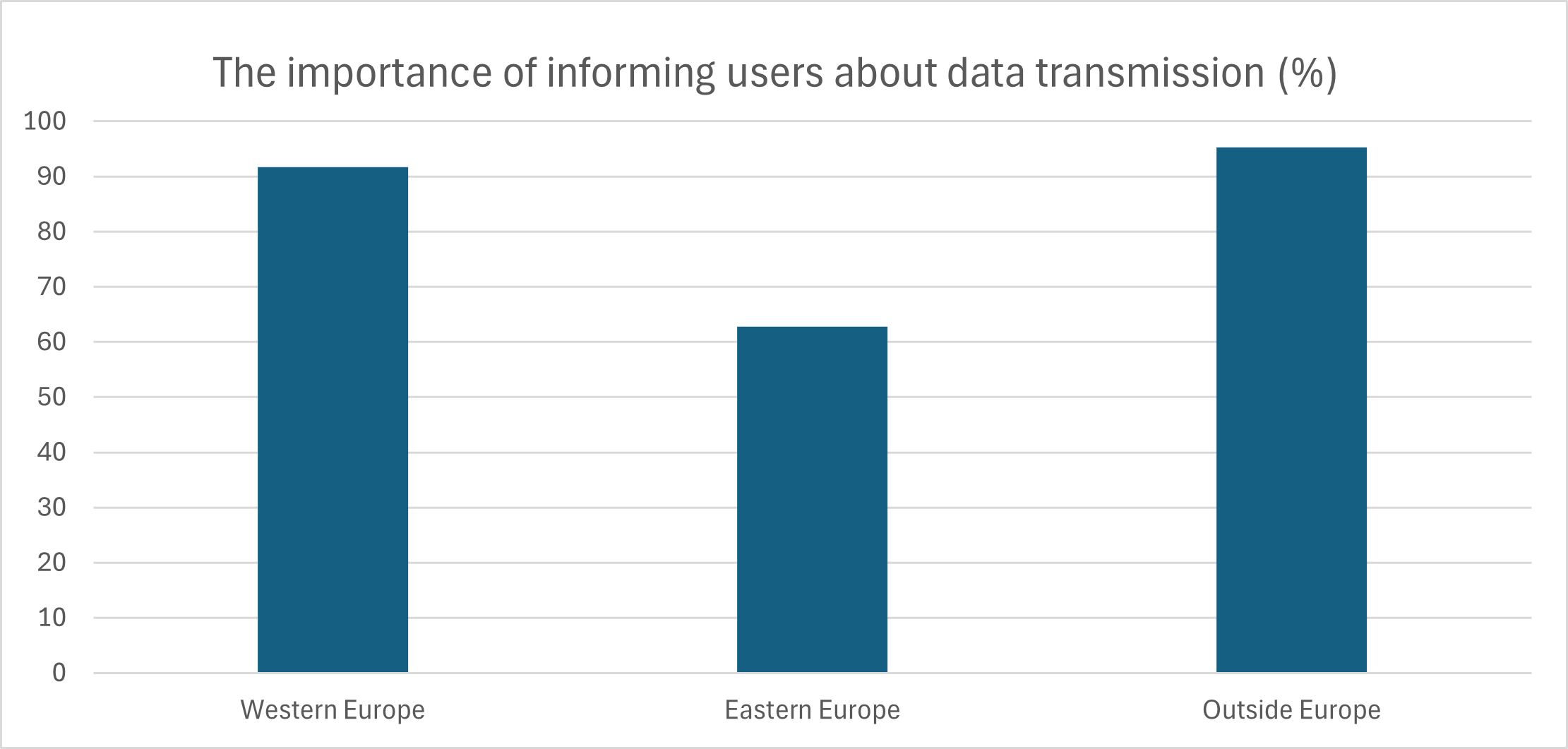}}
\caption{Importance of informing users about data transmission}
\label{fig:importance}
\end{figure}

Interestingly, an overwhelming majority of respondents believe that they should receive clear and detailed information about what kind of data their car collects and where it is sent. Even though there are regional differences within Europe, this expectation is consistently high: in Western Europe, more than 90\% of respondents support this transparency, while in Eastern Europe the proportion is also notably high—above 60\%. In stark contrast, when asked whether they had actually received such information, the vast majority answered negatively, highlighting a significant gap between expectations and current industry practices (Fig. \ref{fig:importance}).

\section{Conclusion}

The proliferation of Internet-connected smart cars marks a significant shift in the automotive landscape. These vehicles, equipped with advanced communication and data processing capabilities, are becoming increasingly common on our roads. However, their growing presence also brings heightened exposure to cybersecurity threats, including remote access vulnerabilities and the potential compromise of sensitive personal data.

Our analysis of 16 international standards and regulations revealed a fragmented and uneven landscape. There is considerable divergence in regulatory strength and enforceability, with some frameworks offering robust guidance while others remain vague or voluntary. The technical depth and specificity of these standards also vary widely, leading to inconsistent treatment of critical issues such as supply chain risks and personal data protection. Notably, no single standard provides comprehensive coverage across all identified threat categories. Each framework reflects the priorities and assumptions of its origin, scope, and target audience, resulting in a patchwork of strengths and blind spots.

Given this complexity, we advocate for a layered approach to securing connected vehicle systems—one that integrates regulatory compliance, technical standards, and organizational best practices. Such a strategy is essential for building systems that are not only resilient to cyber threats but also aligned with legal and ethical expectations.

To complement our regulatory analysis, we conducted a survey to better understand consumer perceptions of smart cars. The results indicate that while most respondents are aware of the potential for their vehicles to transmit data to manufacturers or third parties, they often feel resigned to these risks due to a lack of alternatives. Furthermore, the majority of participants emphasized the importance of being informed about data transmission practices, underscoring a clear demand for greater transparency and user empowerment in the connected car ecosystem.

In summary, the path forward requires both technological vigilance and regulatory evolution. As smart cars continue to evolve, so too must the frameworks that govern their safety, privacy, and trustworthiness.


\vspace{12pt}

\end{document}